# Bayesian probabilistic exploration of Bitcoin informational quanta and interactions under the GITT-VT paradigm

Quan-Hoang Vuong [1,2], Viet-Phuong La [1], Minh-Hoang Nguyen [1]

[1] Centre for Interdisciplinary Social Research, Phenikaa University, Hanoi, Vietnam

[2] Professor, University College, Korea University, Seoul 02841, South Korea.

* **Correspondence:** hoang.nguyenminh@phenikaa-uni.edu.vn

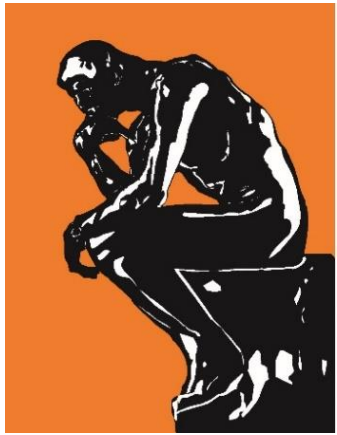

November 20, 2025

[*Original working draft v1*]

## In celebration of Vietnamese Teacher's Day

"— Don't be greedy. Let the greedy fools fight among themselves for a treasure!"

"Luck," *Wild Wise Weird* (2024)

## Abstract

This study explores Bitcoin's value formation through the Granular Interaction Thinking Theory–Value Theory (GITT–VT). Rather than stemming from material utility or cash flows, Bitcoin's value arises from informational attributes and interactions of multiple factors, including cryptographic order, decentralization-enabled autonomy, trust embedded in the consensus mechanism, and socio-narrative coherence that reduce entropy within decentralized value-exchange processes. To empirically assess this perspective, a Bayesian linear model was estimated using daily data from 2022 to 2025, operationalizing four informational value dimensions: Store-of-Value (SOV), Autonomy (AUT), Social-Signal Value (SSV), and Hedonic-Sentiment Value (HSV). Results indicate that only SSV exerts a highly credible positive effect on next-day returns, highlighting the dominant role of high-entropy social information in short-term pricing dynamics. In contrast, SOV and AUT show moderately reliable positive associations, reflecting their roles as low-entropy structural anchors of long-term value. HSV displays no credible predictive effect. The study advances interdisciplinary value theory and demonstrates Bitcoin as a dual-layer entropy-regulating socio-technological ecosystem. The findings offer implications for digital asset valuation, investment education, and future research on entropy dynamics across non-cash-flow digital assets.

**Keywords:** Bitcoin valuation; informational entropy; decentralized socio-technological systems; non-cash-flow digital assets; entropy reduction

## Introduction

Since its inception in 2009, Bitcoin has redefined how value is conceived, transmitted, and collectively sustained. It generates no cash flow, confers no ownership rights, and lacks material utility, yet by 2025, its market capitalization had surpassed USD 2 trillion (IMF, 2025; Wadington, 2025), positioning it among the world's most valuable asset classes. Increasingly regarded as "digital gold" (Baur, Karlsen, Smales, & Trench, 2024; Sygnum Bank, 2021), Bitcoin is primarily valued for its programmed scarcity, resistance to inflation, and perceived stability amid macroeconomic uncertainty.

The persistence of Bitcoin's value presents a substantial challenge to conventional economic reasoning. Unlike traditional assets, its value is not derived from material production, sovereign backing, or industrial utility. Classical and neoclassical economic frameworks situate value within tangible metrics such as labor, production costs, or marginal utility (Damodaran, 2012; Varian, 1996). For example, the Labor Value Theory suggests Bitcoin's worth stems from the effort required to "produce" it (Rotta & Paraná, 2022). Yet, mining is undertaken by algorithmic processes rather than conventional labor, rendering this framework insufficient. Likewise, intrinsic value models are incompatible, as Bitcoin generates neither dividends nor productive cash flows (Abboushi, 2017; Velde, 2013; Williams, 2014). While expected utility theory addresses subjective satisfaction, it fails to justify global consensus over value without a shared, non-negotiable informational substrate (Barberis, Jin, & Wang, 2021; Yadav, 2024). More contemporary approaches—such as behavioral (e.g., prospect theory) and network-based economics—supplement traditional reasoning by incorporating psychological utility, preference formation, and adoption effects. Nevertheless, they still assume that value is anchored in some tangible or quantifiable substrate, whether physical goods, labor exertion, or observable behavioral patterns (Barberis et al., 2021; Soros, 2013; Yadav, 2024).

These theoretical inadequacies arise primarily because disciplinary paradigms conceptualize value in isolation: economists quantify it via price mechanisms; sociologists through normative structures; psychologists through subjective preferences; and physicists through energy and entropy. In the absence of an integrated explanatory mechanism, the notion of "value" remains fragmented across domains. Bitcoin exposes this gap by manifesting a coherent yet non-material value system, demonstrating the need for an interdisciplinary approach to value formation.

To address this gap, we adopt the Granular Interaction Thinking Theory (GITT)'s informational entropy-based value formation (GITT-VT) (Vuong, La, & Nguyen, 2025). GITT is an interdisciplinary information-processing framework integrating principles from quantum mechanics (Hertog, 2023; Rovelli, 2018), information theory (Shannon, 1948), and the mindsponge theory, enabling the simultaneous incorporation of objectivity and subjectivity in value reasoning. Under the GITT-VT perspective, value is not inherent but informational, emerging through probabilistic interactions among informational quanta within a socio-technological system.

Bitcoin's value is therefore better conceptualized as a manifestation of entropy reduction—that is, uncertainty minimization—within decentralized value exchange processes, rather than as a direct function of material-based scarcity or utility. From this perspective, value emerges from informational attributes such as cryptographic order (integrity), trust embedded in the consensus mechanism, and relational coherence across decentralized validation networks. By adopting this lens, the paper advances interdisciplinary value theory and demonstrates how intangible informational systems can generate tangible market outcomes in highly networked, information-driven economies.

To empirically evaluate the plausibility of this theoretical perspective, we investigate how short-term fluctuations in Bitcoin's perceived value are predicted by variables representing informational quanta from the technical subsystem (Store-of-Value – SOV; Autonomy – AUT) and the social subsystem (Social-Signal Value – SSV; Hedonic-Sentiment Value – HSV). This approach enables the disentangling of interactions between low-entropy structural value attributes and high-entropy socio-informational dynamics, facilitating a deeper understanding of Bitcoin's emergent pricing behavior.

## Theoretical Foundation

### *Granular Interaction Thinking Theory-Value Theory (GITT-VT)*

Granular Interaction Thinking Theory (GITT) posits that all observable phenomena emerge from interactions among finite informational quanta (Vuong & Nguyen, 2024a, 2024b). Informed by quantum mechanics and information theory, GITT conceptualizes reality as fundamentally granular, relational, and indeterminate:

- a) Granularity: Information exists in discrete, finite units; cognitive and socio-economic systems can process only limited "grains" of information at any given time.
- b) Relationality: Value does not reside within information units themselves but emerges from their interactions within and across systems.
- c) Indeterminacy: Outcomes are probabilistic, meaning that identical informational inputs may lead to different valuations depending on contextual variables and prior informational states.

According to GITT, every cognitive or socio-economic system processes information to minimize entropy, seeking to maximize coherence while conserving energy (Vuong, La, et al., 2025). Within this paradigm, value is not inherent but emergent. At the individual level, value emerges when information—such as lived experiences, perceptions, beliefs, biological predispositions, emotions, and worldviews—interacts with newly absorbed information from environmental, socio-cultural, and economic contexts to produce a lower-entropy mental state. This state supports the individual's survival, growth, and reproduction, while conserving cognitive and physiological energy. At the societal level, value emerges through interactions among informational components within the social system—such as historical experiences, collective beliefs, and scientific knowledge—with newly acquired information, including technological innovations, policy signals, and ecological shifts. These interactions foster system-wide coherence and adaptability, reducing socio-economic entropy and helping society manage uncertainty while addressing collective needs.

In essence, value represents the information retained with the highest probability within a system because of its relevance to existence and stability. Drawing on Shannon (1948)'s entropy formula, informational entropy $H(X_t)$ of a system $X$ is calculated as follows:

$$H(X_t) = -\sum P_i(t) \log_2 P_i(t)$$

Here, $H(X)$ denotes the informational entropy of a random variable $X$ with possible outcomes $\{x_1, x_2, \ldots, x_n\}$ and corresponding probabilities $\{P(x_1), P(x_2), \ldots, P(x_n)\}$. Each probability $P(x_i)$ expresses how likely the outcome $x_i$ is to occur. According to this framework, entropy rises sharply when informational units accumulate without clear prioritization, reaching its maximum when all information is equally weighted—that is, when $P(x_i) = \frac{1}{n}$. Thus, a decrease in $H(X_t)$, either by allocating probability to specific information or by reducing the number of pieces of information within the system, signifies an increased order, hence higher perceived values.

### *Bitcoin as an Entropy-Regulating Socio-technological Ecosystem*

In socio-economic systems, informational quanta continuously compete for attention, retention, and interpretive dominance. Only information that reliably reduces uncertainty—by demonstrating credibility, coherence, or simplicity—is retained and gradually internalized as value quanta. These value quanta stabilize cognitive and market expectations by reinforcing belief coherence and reducing entropy. When such information recurs frequently and aligns with pre-existing belief structures, it shapes shared valuation logics that transcend material substance. Bitcoin exemplifies this mechanism. Its valuation is not derived from intrinsic utility (Umlauft, 2018), but from the systemic regulation of informational ambiguity across two interacting subsystems: a technical entropy system and a social entropy system (see Figure 1).

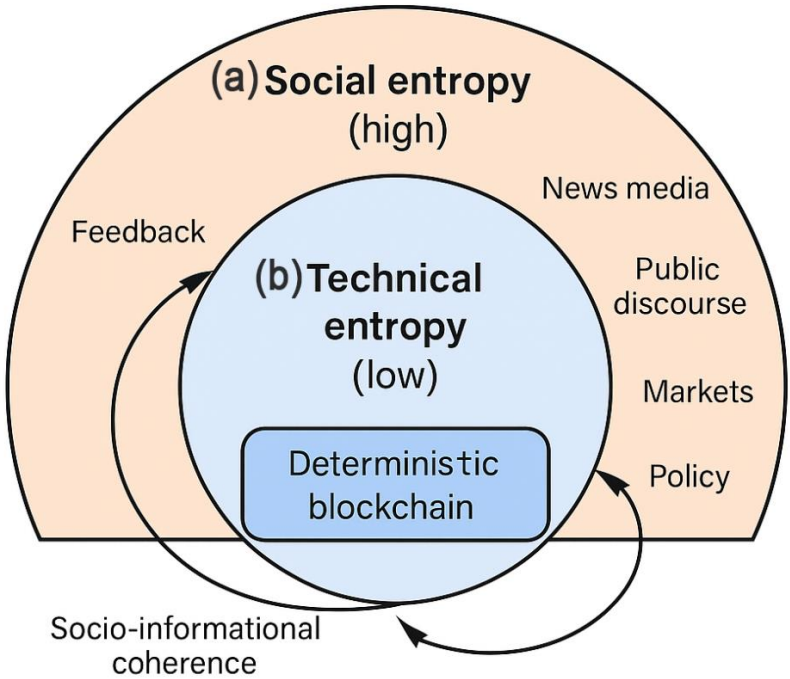

**Figure 1:** The Bitcoin's Entropy Regulating Socio-Technological Ecosystem

Bitcoin's ability to maintain low technical entropy is primarily driven by three complementary characteristics: deterministic protocol architecture, cryptographic integrity, and informational scarcity. First, its blockchain architecture operates under deterministic rules, meaning that each recorded transaction and block validation follows predefined algorithms without ambiguity. This eliminates stochastic variability in the system's core mechanics and ensures that every action on the network is traceable, reproducible, and algorithmically verifiable. Second, Bitcoin employs robust cryptographic mechanisms, such as SHA-256 hashing and Proof-of-Work consensus, which regulate how transactions are confirmed and prevent retroactive manipulation. The inherent computational irreversibility of these cryptographic functions makes past data technically immutable, reinforcing informational order by prohibiting entropy-generating alterations.

A third key feature is Bitcoin's informational scarcity, which fundamentally differs from the material scarcity of traditional assets. The 21-million-coin limit, encoded directly into the protocol, represents a fixed boundary condition that is both predictable and universally verifiable across nodes. This ensures that supply-side uncertainty—one of the most common sources of entropy in traditional financial assets—is structurally eliminated. Additionally, each newly mined block functions as a discrete informational quantum that contributes to entropy stabilization. Through consensus, uncertain network states (e.g., unverified transactions) are progressively converted into orderly, immutable ledger entries. As more nodes replicate the exact ledger globally—introducing redundancy in Shannon's sense—the probability of informational inconsistency diminishes. Thus, Bitcoin maintains low systemic entropy through self-stabilizing, algorithmically governed information production and distribution, making technical reliability a foundational element of its perceived economic value.

Grounded on Bitcoin's low-entropy technical subsystem, the social subsystem operates within a high-entropy environment influenced by market speculation, media narratives, regulatory uncertainty, and community discourse. While the technical layer offers structural determinism through cryptographic verification and immutable protocol rules, the social layer reflects fluidity and unpredictability. Informational shocks—such as security breaches, geopolitical shifts, or sudden policy interventions—temporarily amplify entropy by destabilizing collective expectations. In response, the system relies on feedback mechanisms, whereby corrective communication, public transparency, and data-backed assurances reinterpret disorder as order, facilitating entropy dissipation. The efficiency with which the system converts informational ambiguity into coherent signals determines the degree of valuation stability. Accordingly, Bitcoin's market value is not merely the product of speculative demand; rather, it reflects the system's capacity to maintain entropy equilibrium through recursive interplay between technological robustness and socio-informational stabilization.

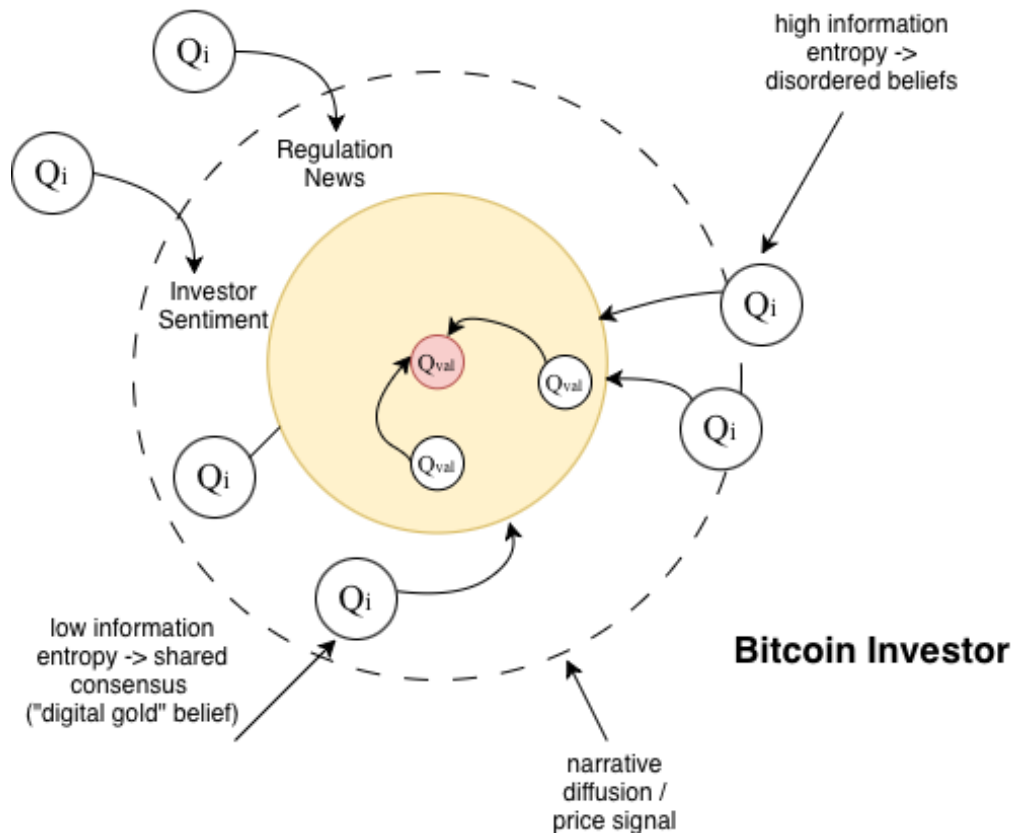

**Figure 2:** Information-Value Interaction Diagram

The conceptual diagram of these interaction processes is visualized in Figure 2. Specifically, informational quanta ($Q_i$) from high-entropy environments continuously enter cognitive and market systems and are filtered through bitcoin investors' mindsponge mechanisms. Those interactions that effectively reduce informational entropy are internalized as value quanta ($Q_{val}$), manifesting as perceived economic worth. Feedback loops then reintroduce these low-entropy informational structures into the infosphere—often crystallized as coherent narratives (e.g., "digital gold")—further reinforcing belief stability and reducing systemic uncertainty. The dynamic interactions between Bitcoin's technical and social entropy subsystems reflect the three foundational principles of GITT: granularity, relationality, and indeterminacy.

- **Granularity:** Each Bitcoin transaction, validated block, or communicative event constitutes a discrete informational grain. This finiteness allows uncertainty to be progressively quantified and reduced. Consistent with Shannon (1948)'s redundancy principle, the replication of identical ledger records across distributed nodes enhances informational reliability, lowering transmission entropy. Granularity also manifests socially through individual tweets, policy announcements, price reports, and media commentaries, which act as micro-level informational quanta that cumulatively influence macro-level market valuation dynamics.

- **Relationality:** The value of Bitcoin does not arise from isolated informational inputs but emerges from the interaction between technical assurance (e.g., cryptographic confirmations), cognitive status (e.g., investor confidence, knowledge, fear, excitement), and social reinforcement (e.g., media coverage, expert commentary, influential endorsement). When high-credibility inputs—such as central bank recognition or ETF approvals—resonate with pre-existing belief quanta (e.g., "Bitcoin is sound monetary infrastructure"), informational entropy declines through the reinforcement of coherent valuation narratives. Conversely, discordant quanta—such as regulatory crackdowns or exploit attempts—increase entropy, inducing market volatility. Thus, Bitcoin's valuation trajectory can be conceptualized as a continuous oscillation between entropy reduction (trust consolidation) and entropy expansion (informational shock).

- Indeterminacy: GITT's indeterminacy principle posits that future value states are inherently probabilistic rather than deterministic. Bitcoin's price $V$, therefore, reflects a probability distribution of belief coherence among market participants. Each new informational input $I_t$ updates this distribution according to Bayesian inference, as expressed below (Vuong, Nguyen, Ho, & La, 2025):

$$P(V \mid I_t) = \frac{P(I_t \mid V)P(V)}{\sum P(I_t \mid V_i)P(V_i)}$$

  Where $P(V \mid I_t)$ represents the updated belief in Bitcoin's value upon receiving new information $I_t$. Entropy decreases as informational coherence and reliability increase, and vice versa.

  When new information—such as evidence of protocol resilience or institutional adoption—is consistent with entrenched value quanta, posterior belief probability increases, reducing entropy. Over time, recursive updating yields path-dependent stabilization of collective valuation, illustrating how intangible informational processes can translate into tangible economic outcomes.

In general, Bitcoin's value formation can be conceptualized as a three-stage entropy-regulation process within the socio-technological ecosystem (Vuong & Nguyen, 2024b).

First, information proliferation—driven by media diffusion, market speculation, and socio-technological developments—introduces large quantities of heterogeneous data into the system, increasing informational entropy through uncertainty and interpretive divergence. After that, through cognitive and socio-informational filtering mechanisms, alongside technical validation systems (e.g., blockchain consensus, cryptographic verification, distributed ledger replication), the system selectively absorbs and processes incoming information. This drives narrative convergence and data convergence, simultaneously reducing entropy through alignment of belief structures and confirmation of transaction authenticity. Third, as coherent narratives and validated technical data are consistently reinforced by the system's reliability and social endorsement, they converge into low-entropy informational states that are retained collectively and perceived as economic value. Bitcoin's endurance across market cycles exemplifies this entropy-regulation process. Despite volatility and episodic entropy surges, its core informational and technical architecture remains intact, enabling recursive reinforcement of transactional trust and valuation stability.

To empirically assess the plausibility of this understanding, we analyze how short-term fluctuations in Bitcoin's perceived market value are predicted by variables operationalizing informational quanta from both its technical subsystem—represented by Store-of-Value (SOV) and Autonomy (AUT)—and its social subsystem—captured through Social-Signal Value (SSV) and Hedonic-Sentiment Value (HSV). This dual-layer analytical design is consistent with the GITT–VT framework, which posits that value arises from interactions between low-entropy structural attributes (e.g., cryptographic reliability, scarcity, decentralization) and high-entropy socio-informational dynamics (e.g., narrative-driven attention, memetic transmission, sentiment shifts). By incorporating both types of informational quanta within a time-series Bayesian modeling, we aim to disentangle the respective roles and interaction mechanics of stable, long-term value anchors and transient, high-velocity informational shocks. This enables a more granular understanding of Bitcoin's emergent pricing behavior and provides empirical evidence for value as an emergent phenomenon of entropy regulation within a decentralized socio-technological ecosystem.

# Methodology

## *Data Sources and Variables*

This study adopts a quantitative research design to investigate which perceived value dimensions' fluctuations influence short-run price dynamics of Bitcoin. Grounded in the GITT-VT, we conceptualize Bitcoin's valuation as multidimensional and operationalize it through four information-derived value constructs: Store-of-Value (SOV), Autonomy (AUT), Social-Signal Value (SSV), and Hedonic/Experiential Value (HSV). These dimensions are treated as informational quanta reflecting distinct cognitive and socio-technological drivers and are incorporated into a Bayesian regression model to assess their explanatory power on Bitcoin's daily returns.

SOV and AUT represent low-entropy structural components, anchoring valuation through beliefs in scarcity, decentralization, and functional reliability. In contrast, SSV and HSV represent high-entropy socio-cognitive components, shaped by market sentiment, narrative intensity, social discourse, and emotional engagement. This modeling approach enables the empirical examination of reflexive feedback loops, wherein changes in perceived value influence price movements, which subsequently reshape perception through iterative belief updating.

Daily time-series data were collected from December 2022 to November 2025 using multiple reputable sources to operationalize Bitcoin's multidimensional perceived value within the GITT-VT framework. Table 1 outlines the operational proxies corresponding to each value dimension, the data source, and the measurement description.

**Table 1:** Variable description

| Variables | Description | Operational Proxy | Source | Measurement Description |
|---|---|---|---|---|
| SOV | Store-of-Value | Exchange Supply Ratio (ESR) | Glassnode | Ratio of BTC held on exchanges to total circulating supply. The inverse (1–ESR) reflects long-term holding behavior |

| | | | | and scarcity perception. |
|---|---|---|---|---|
| AUT | Autonomy | Active Addresses | Cryptoquant | Number of daily Active Addresses reflects the extent to which Bitcoin is used independently of intermediaries - an indicator of autonomy-based value. |
| SSV | Social-Signal Value | Google Trends Index | Google Trends | Measures collective attention and narrative intensity toward "Bitcoin". |
| HSV | Hedonic Value | Fear & Greed Index | Alternative.me | Quantifies investor sentiment, capturing emotional oscillations between fear and greed. |
| BTC_Return | Bitcoin's daily return | Daily log return | Glassnode | $Return_t = \ln(Price_t/Price_{t-1})$ |

All variables were aligned to a daily frequency to ensure temporal consistency and standardized prior to estimation to facilitate coefficient interpretability across differing value scales. This dataset allows examination of how changes in perceived informational value influence Bitcoin's price dynamics within the entropy-regulation framework.

Data were preprocessed using the following procedures:

- **Normalization:** All proxy variables were rescaled to a 0–100 range to harmonize measurement units and mitigate disparities in magnitude across indicators.

- **Differencing:** To isolate perception-driven shocks rather than structural levels, first differences were computed using

$$\Delta X_t = X_t - X_{t-1}$$

- **Lag Structure:** A one-day lag ($\Delta X_{t-1}$) was introduced to mitigate simultaneity bias between changes in perceived value dimensions and price responses.

- **Stationarity Assessment:**

    - Augmented Dickey–Fuller (ADF) tests confirmed stationarity for all first-order differenced variables.

    - Descriptive statistics verified data completeness with no missing observations following temporal alignment.

- **Time-Range Consistency:** All proxies were synchronized on a daily frequency, yielding approximately N ≈ 1,090 observations over the study period.

*Statistical Model*

The empirical analysis employs a Bayesian linear regression model to assess how daily shifts in perceived value dimensions affect short-run Bitcoin returns. The model is specified as follows:

$$BTC\_Return_t \sim Normal(\mu_t, \sigma)$$

$$\mu_t = \alpha + \beta_1 \Delta SOV_{t-1} + \beta_2 \Delta AUT_{t-1} + \beta_1 \Delta SSV_{t-1} + \beta_1 \Delta HSV_{t-1}$$

$$\beta \sim Normal(M, S)$$

where $Return_t$ denotes the daily logarithmic return of Bitcoin, and $\Delta X_{t-1}$ represents the lagged first difference of each value perception proxy. The inclusion of a one-day lag aims to reduce simultaneity bias by ensuring temporal precedence between perception change and price reaction. The residual variance is captured by $\sigma$, assumed to be constant over time. The coefficient values follow a normal distribution with a mean denoted by $M$ and a standard deviation denoted by $S$. The logical network of the model is visualized in Figure 3.

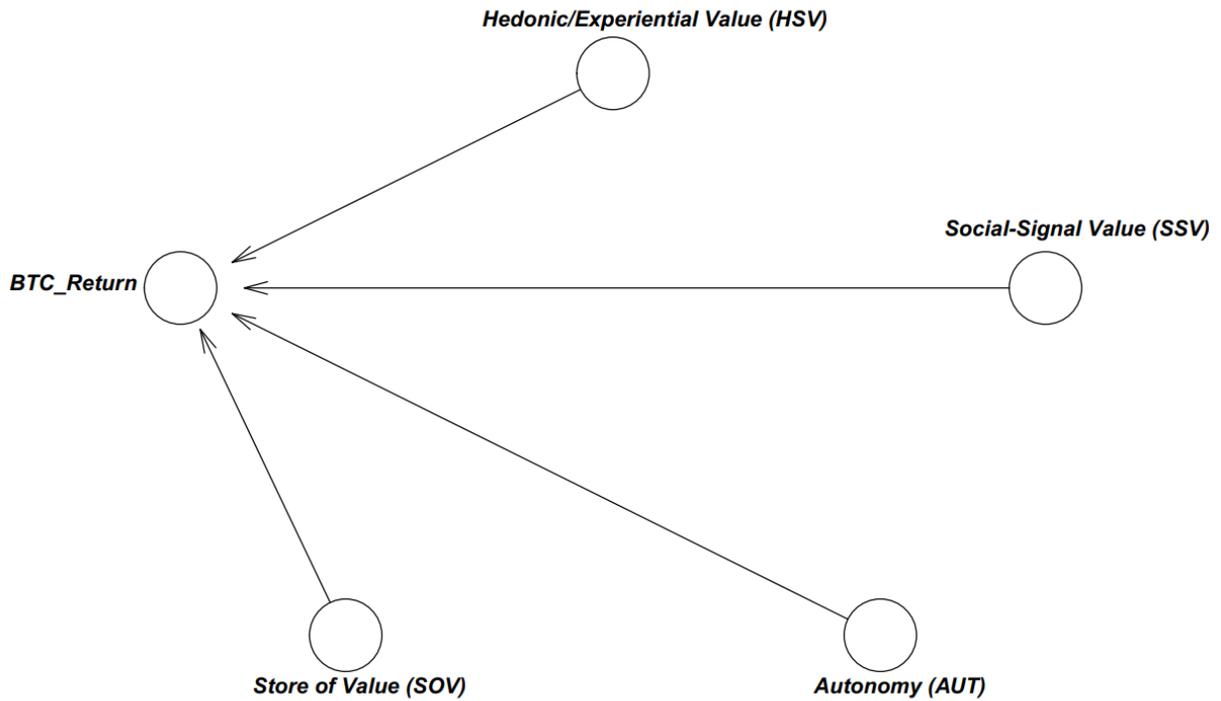

**Figure 3:** Model's logical network

To ensure stable estimation while allowing sufficient flexibility, weakly informative priors were applied:

$$\alpha \sim N(0,1)$$

$$\beta_i \sim N(0,1) \text{ for } i = 1, \dots, 4$$

$$\sigma \sim Cauchy(0,1)$$

These priors provide mild regularization, reduce the risk of overfitting, and enable the observed data to drive inference. Parameter estimation was conducted via Markov Chain Monte Carlo (MCMC) sampling using four chains with 4,000 iterations each.

***Analysis and Validation***

We applied the Bayesian Mindsponge Framework (BMF) analytics to test the proposed hypotheses. This framework combines the theoretical reasoning strength of GITT-VT with the inferential power of Bayesian analysis, providing a robust tool for examining cognitive, psychological, and social dynamics (Vuong, Nguyen, & La, 2022). GITT is inherently probabilistic, positing that improvements in cognitive and social processes emerge from the continuous interaction between existing cognitive structures and newly absorbed information. This probabilistic mechanism aligns naturally with the Bayesian approach, which treats both known and unknown entities—such as unobserved data, uncertainties, and model parameters—as probabilistic variables (Gill, 2015).

Bayesian inference offers several advantages over conventional frequentist methods. The limitations of frequentist statistics have been widely cited as contributors to the reproducibility crisis in socio-psychological research (Camerer et al., 2018; Open Science Collaboration, 2015). A core issue is the instability and frequent misinterpretation of *p*-values (Halsey, Curran-Everett, Vowler, & Drummond, 2015). Originally designed to evaluate evidence against the null hypothesis, *p*-values are often used as rigid cut-off points (commonly 0.05) to determine "significance," despite their high variability even under typical power conditions (80%) (Halsey et al., 2015). To address this problem, several authors have proposed estimation-based approaches and graphical interpretation using means and confidence intervals as more informative alternatives (Masson & Loftus, 2003).

Bayesian inference provides a conceptually coherent alternative by integrating estimation and visualization within a unified probabilistic structure. Unlike frequentist confidence intervals—which reflect that across repeated sampling, 95% of such intervals would contain the true parameter—a 95% Bayesian credible interval specifies the probability that the parameter lies within the given range, conditional on the observed data and model assumptions (van Zyl, 2018; Wagenmakers et al., 2018). Bayesian analysis also enhances transparency, interpretability, and suitability for contemporary socio-behavioral research, where uncertainty and contextual complexity are central to reasoning.

Another key benefit of Bayesian inference is its ability to integrate informative priors, thereby mitigating multicollinearity—an issue commonly observed in health-related studies. As Leamer (1973) argued that multicollinearity often represents a "weak data problem," in which large standard errors lead to unstable estimates. In Bayesian terms, this is reflected in overlapping prior and posterior distributions within certain parameter spaces. Incorporating informative priors can constrain estimates, making them more stable and robust (Adepoju & Ojo, 2018; Jaya, Tantular, & Andriyana, 2019). However, concerns regarding subjectivity in prior specification are valid. To address this, the primary analyses in this study used weakly informative priors to minimize prior influence on the results.

After model estimation, we employed Pareto-smoothed importance sampling leave-one-out cross-validation (PSIS-LOO) diagnostics (Vehtari & Gabry, 2024; Vehtari, Gelman, & Gabry, 2017) and posterior predictive checks (Gelman & Meng, 1996) to examine whether the fitted model is compatible with the observed data. First, the LOO criterion was calculated using:

$$LOO = -2LPPD_{loo} = -2\sum_{i=1}^{n} \log \int p(y_i|\theta)p_{post(-i)}(\theta)d\theta$$

In this expression, $p_{post(-i)}(\theta)$ represents the posterior distribution estimated without the influence of observation $i$. The PSIS method provides $k$-Pareto diagnostics to identify potentially influential observations: values below 0.5 indicate a good model fit, whereas values above 0.7 may suggest influential data points that could bias the LOO estimate.

Posterior predictive checks were then conducted to evaluate model adequacy by comparing simulated predictions with the observed data. The posterior predictive distribution is defined as:

$$p(y^{rep}|y) = \int p(y^{rep}|\theta)p(\theta|y)d\theta,$$

where $y$ denotes the observed data, $\theta$ is the vector of estimated parameters, and $y^{rep}$ represents the replicated (model-predicted) data. A model is considered to fit well if the distribution of the replicated data closely aligns with that of the observed data, typically assessed through visual comparison. When this alignment is evident, it suggests that the model adequately captures the observed data.

Following model fit evaluation, convergence of the Markov chains was examined using both statistical and visual diagnostics. Effective sample size (*n_eff*) reflects sample sufficiency, with values exceeding 1,000 indicating reliable estimates (Mallik, 2021). The Gelman–Rubin statistic (*Rhat*) was used to assess convergence, where values approaching 1 denote satisfactory mixing of chains (Brooks & Gelman, 1998). Additionally, trace plots, Gelman-Rubin-Brooks plots, and autocorrelation plots were visually examined to confirm Markov chain convergence further.

Bayesian analyses were implemented in R using the bayesvl package version 1.0.0 (La & Vuong, 2019; Vuong & La, 2025). To promote transparency and reproducibility, the complete dataset and analytical code have been openly deposited on Zenodo and are freely accessible at https://zenodo.org/records/17657145, enabling independent replication and further scholarly use.

## Results

### *Model Fit and Convergence Diagnostics*

To examine the short-term price effects of shifts in Bitcoin's perceived value dimensions, a Bayesian regression model was estimated using one-day lagged differences of four value components: Store-of-Value (SOV), Autonomy (AUT), Social-Signal Value (SSV), and Hedonic Value (HSV). Daily logarithmic returns were modeled under weakly informative priors to promote data-driven inference while maintaining parameter stability. Posterior estimation was performed using Markov Chain Monte Carlo (MCMC) simulation.

Prior to interpreting the regression results, model fit was evaluated. According to established guidelines, *k*-values below 0.7 indicate acceptable predictive reliability. As illustrated in Figure 4, all estimated *k*-values were well below 0.5, demonstrating a strong correspondence between the model and the observed data.

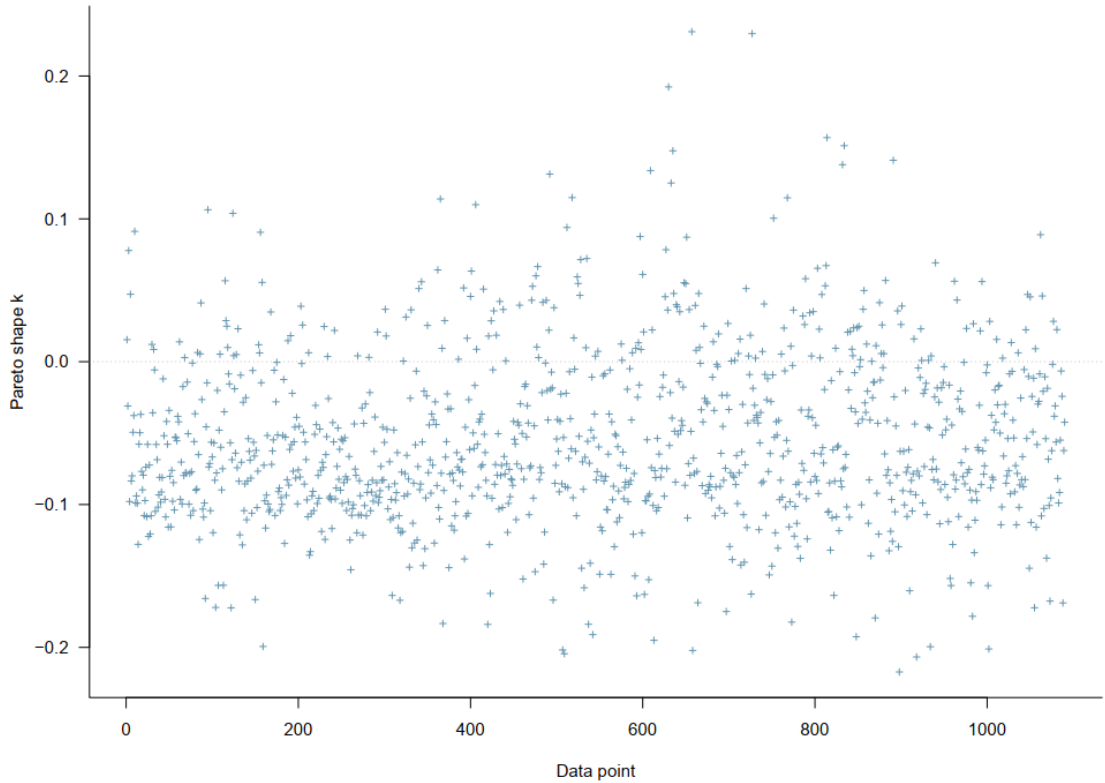

**Figure 4:** PSIS-LOO diagnostic plot

The posterior predictive check further confirmed the model's adequacy. In Figure 5, the dark line represents the distribution of the observed outcomes ($y$), while each of the 100 lighter lines corresponds to the kernel density estimate of a replicated dataset ($y^{rep}$) drawn from the posterior predictive distribution. The close alignment between the observed distribution and the replicated distributions indicates that the model captures the underlying data-generating process well, thereby supporting its validity.

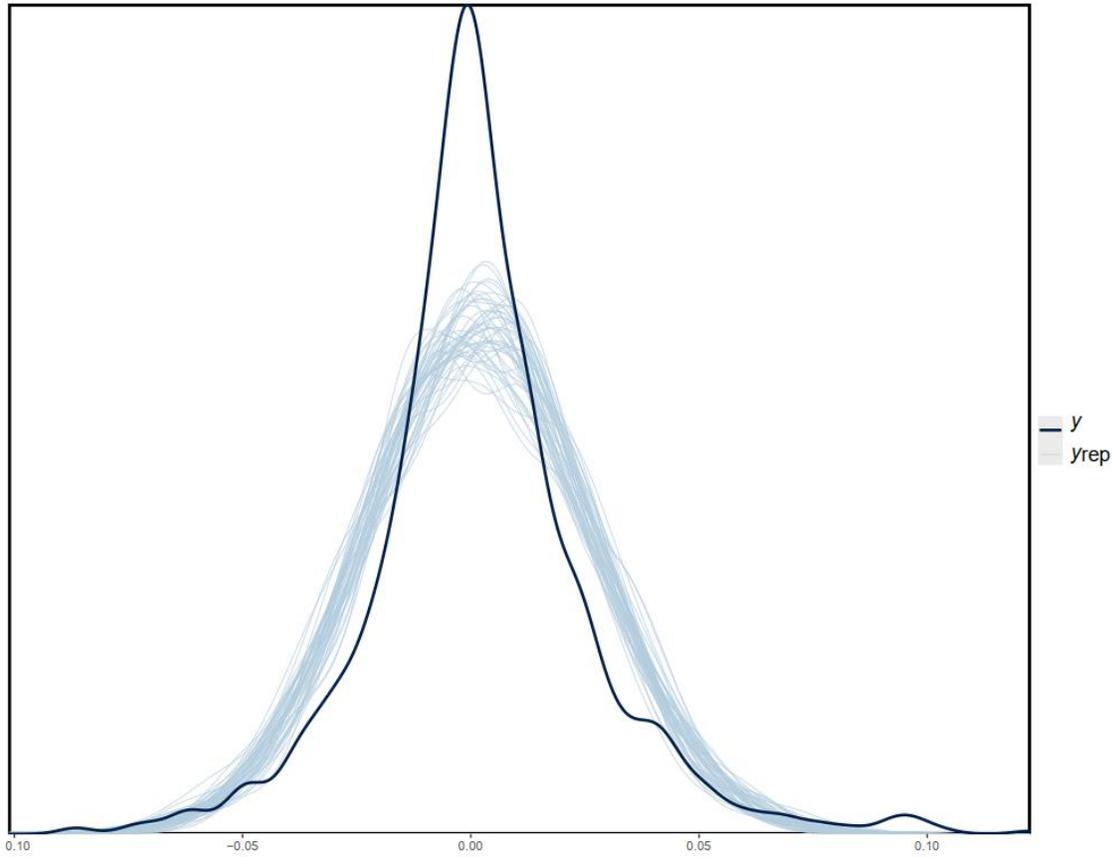

**Figure 5:** Graphical posterior predictive check

Posterior summaries of all parameters—including means, standard deviations, 95% Highest Posterior Density Intervals (HPDIs), effective sample sizes (*n_eff*), and Gelman–Rubin convergence statistics (*Rhat*)—are presented in Table 2. All parameters exhibit *Rhat* = 1.00, confirming full convergence, while *n_eff* exceeds 2,000, indicating highly efficient chain mixing and robust posterior estimation.

In addition, the convergence of the Markov chains was further assessed using trace plots, Gelman–Rubin–Brooks plots, and autocorrelation diagnostics. The trace plots (Figure 4) display the simulated MCMC values across successive iterations, with the x-axis representing iteration number and the y-axis representing the parameter estimates. The chains exhibit good mixing and remain stationary around an equilibrium, indicating convergence.

Trace plots, Gelman-Rubin-Brooks plots, and autocorrelation plots are also used to validate Markov chain convergence. The trace plots in Figure 6 illustrate the MCMC sample values after each successive iteration along the chain. The *y*-axis demonstrates the coefficient's value, while the *x*-axis reflects the number of iterations of the Markov process. As the chains are good-mixing and stationary around an equilibrium, the Markov chains are convergent.

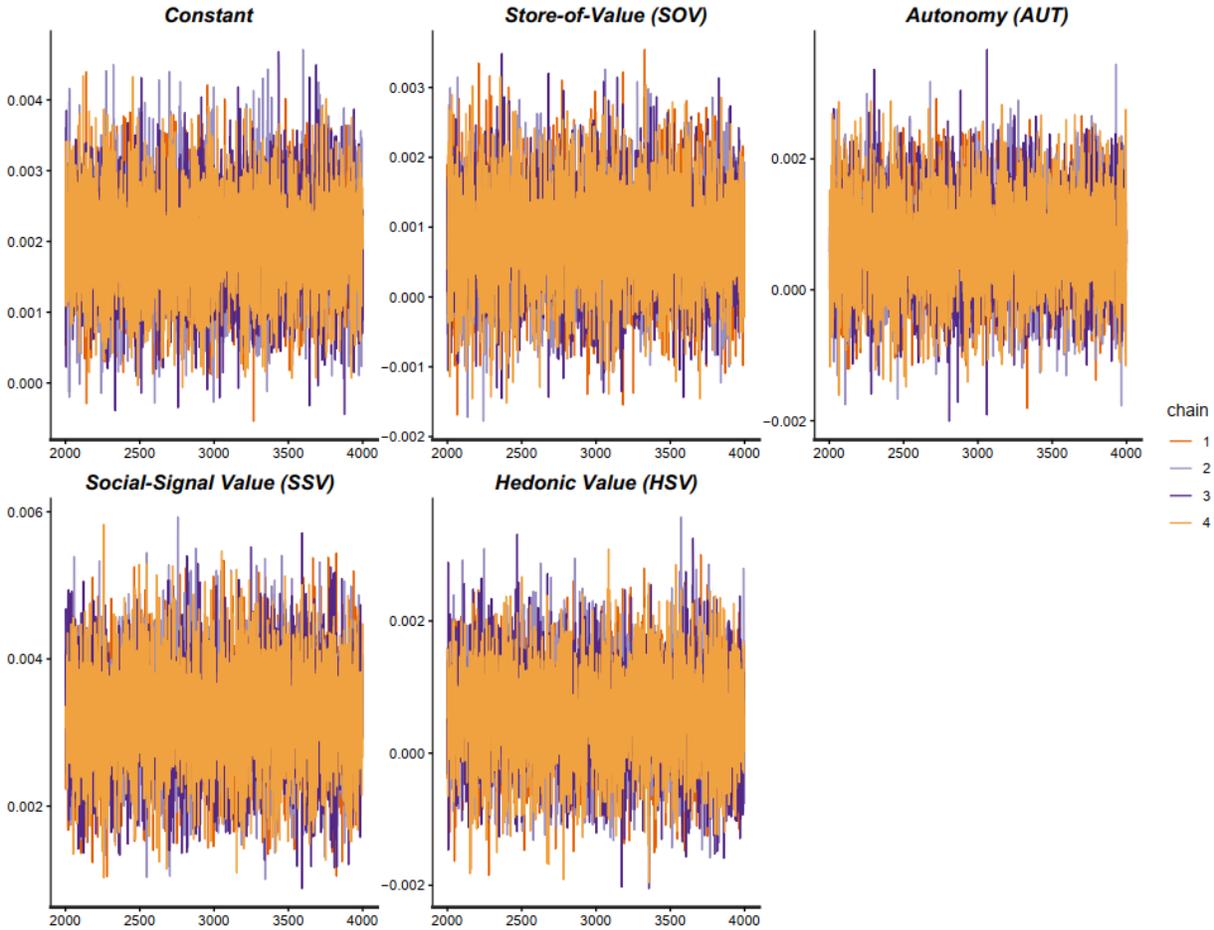

**Figure 6:** Model's trace plots

Similarly, the Gelman–Rubin–Brooks diagnostics (Figure 7) confirm convergence. The Gelman–Rubin statistic (*Rhat*) indicates sampling efficiency, and values approaching 1 indicate adequate convergence across chains. As shown in Figure 7, all *Rhat* values drop rapidly to 1 before the completion of the warm-up period, indicating satisfactory convergence.

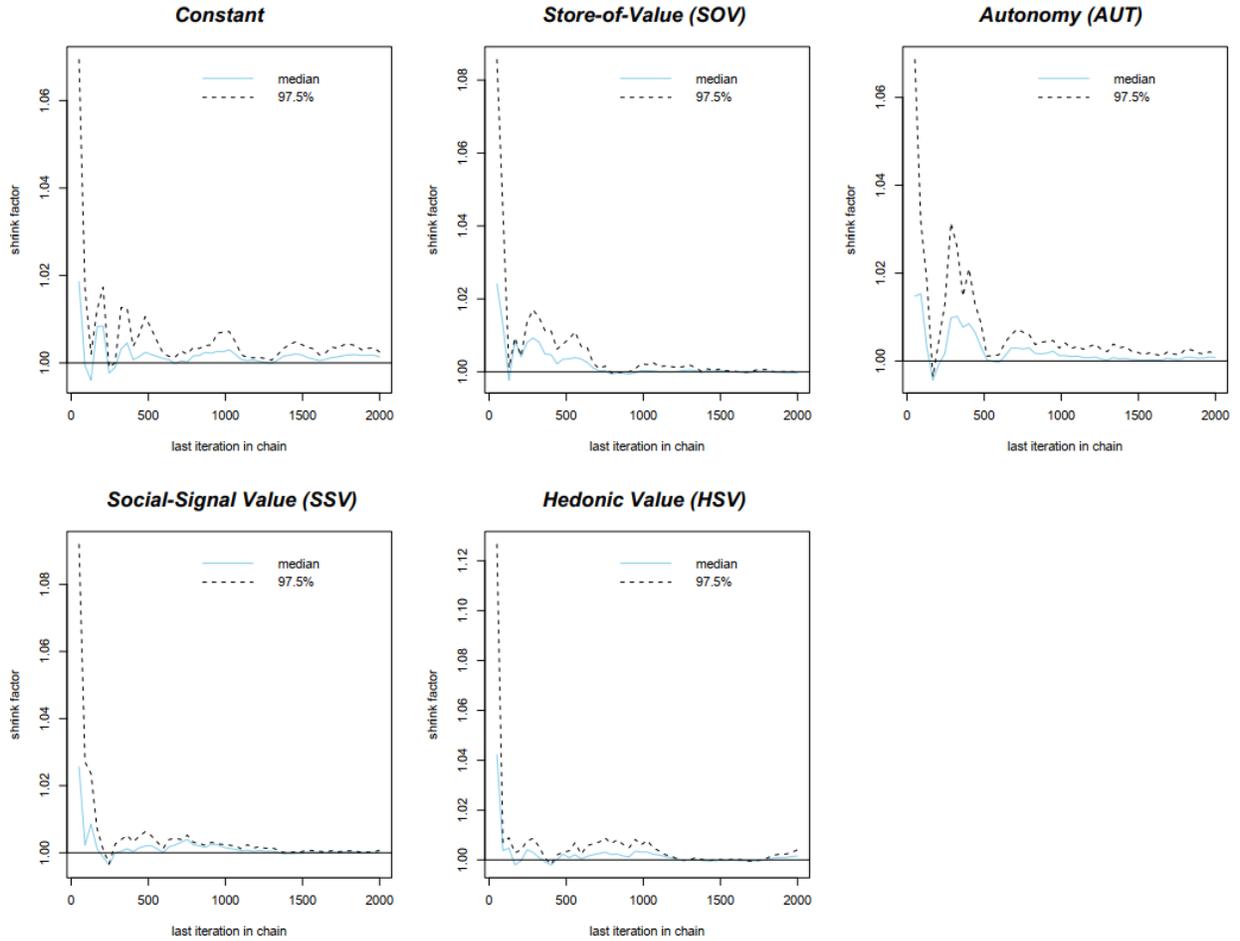

**Figure 7:** Model's Gelman-Rubin-Brooks plots

Autocorrelation plots were also used to verify the Markov property of the sampling process. These plots depict the degree of correlation between samples at increasing lag intervals (e.g., lag 0 representing self-correlation, lag 1 representing correlation with the subsequent sample, and so on). For estimates to be unbiased, MCMC samples should be approximately independent. As illustrated in Figure 8, autocorrelation decreases rapidly, indicating that samples are effectively uncorrelated. This suggests that the chains are memoryless and validates the stochastic simulation process.

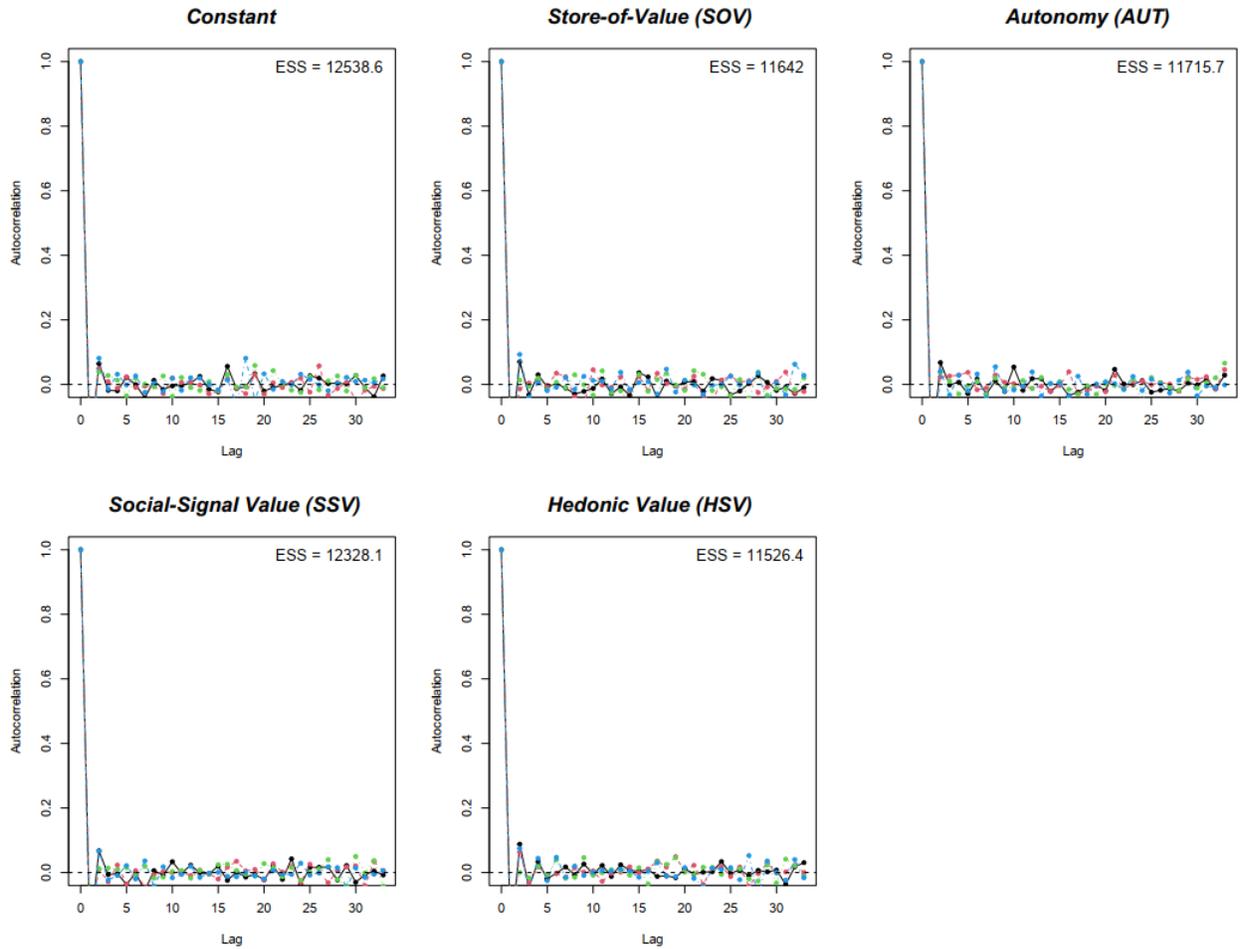

**Figure 8:** Model's autocorrelation plots

*Posterior Distributions*

As seen in Table 2, the intercept has a posterior mean of 0.00199 and a 95% HPDI of [0.00057, 0.00340], suggesting that Bitcoin exhibits a slight positive drift on days with no substantial perception shocks, with 95% credibility. This implies a minor but persistent upward bias in returns under stable informational conditions.

**Table 2:** Estimated posterior distributions

| Parameters | Mean | Standard Deviation | 2.50% | 97.50% | n_eff | Rhat |
|---|---|---|---|---|---|---|
| | | | | | | |
| *Constant* | 0.00199 | 0.00073 | 0.00057 | 0.00340 | 11611 | 1 |
| Store-of-Value ($\Delta SOV_{t-1}$) | 0.00084 | 0.00074 | -0.00062 | 0.00231 | 11254 | 1 |
| Autonomy ($\Delta AUT_{t-1}$) | 0.00073 | 0.00073 | -0.00073 | 0.00219 | 11247 | 1 |
| Social-Signal Value ($\Delta SSV_{t-1}$) | 0.00324 | 0.00072 | 0.00183 | 0.00465 | 11823 | 1 |
| Hedonic/Experiential Value ($\Delta HSV_{t-1}$) | 0.00058 | 0.00073 | -0.00087 | 0.00202 | 11128 | 1 |

The lagged changes in Store-of-Value ($\Delta SOV_{t-1}$), Autonomy ($\Delta AUT_{t-1}$), and Social-Signal Value ($\Delta SSV_{t-1}$) tend to be positively associated with next-day Bitcoin returns. However, as illustrated in Figure 9, the 95% Highest Posterior Density Intervals (HPDIs) for Store-of-Value [-0.00062,0.00231] and Autonomy [-0.00073,0.00219] partially extend into the negative range. Thus, to gauge the reliability of these associations, both mean estimates and standard deviations were considered. Because the posterior mean of Store-of-Value ($\Delta SOV_{t-1}$) slightly exceeds its standard deviation, and the posterior mean of Autonomy ($\Delta AUT_{t-1}$) is approximately equal to its standard deviation, their positive effects can be regarded as moderately reliable.

In contrast, the 95% HPDI of Social-Signal Value ($\Delta SSV_{t-1}$) lies entirely above zero [0.00183,0.00465], indicating a robust and highly credible positive association between the lagged change in Social-Signal Value and next-day Bitcoin returns. Conversely, the lagged shift in Hedonic/Experiential Value ($\Delta HSV_{t-1}$) exhibits a posterior mean close to zero and a tightly bounded HPDI centered around zero, implying no meaningful or credible impact on next-day returns.

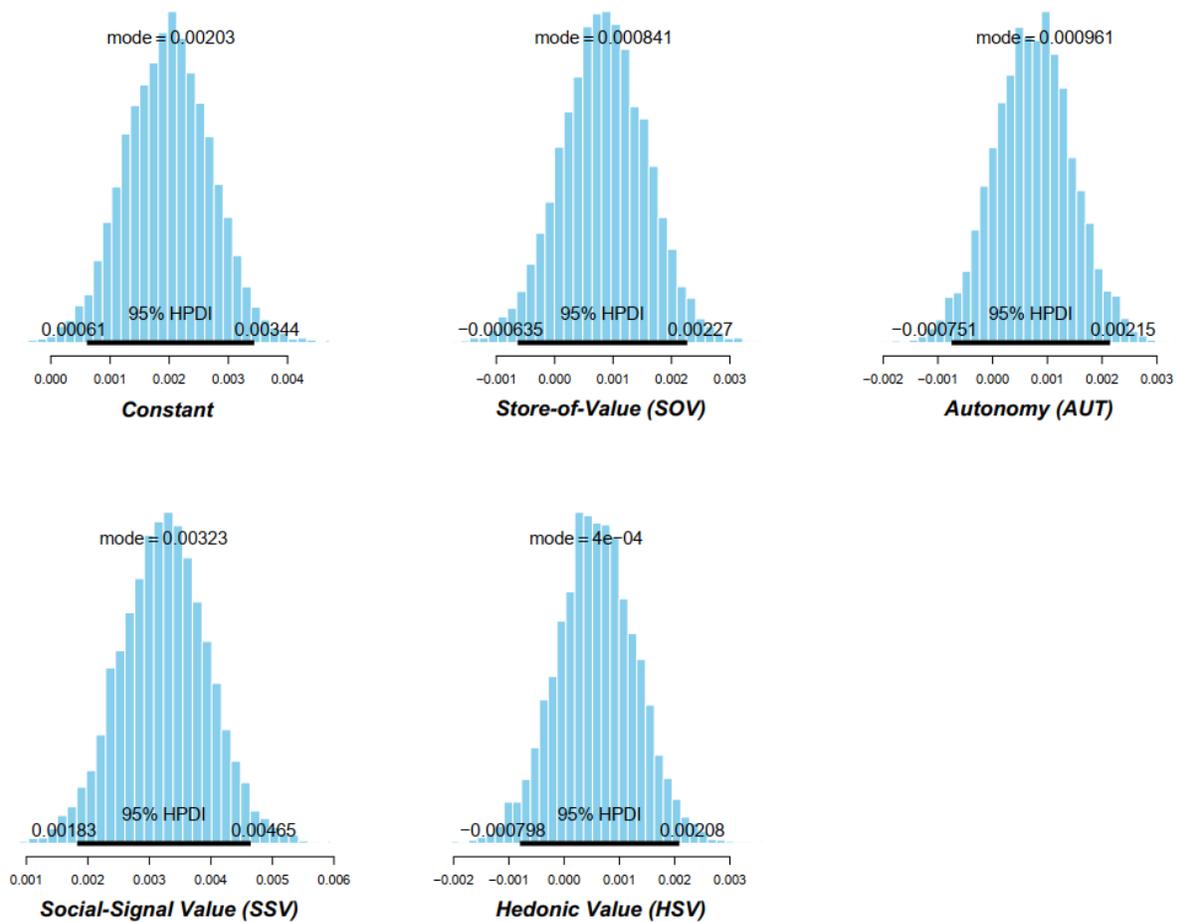

**Figure 9:** Model's posterior distributions

## Discussion

The present study applies the GITT–VT framework to explore Bitcoin's value formation and conceptualizes Bitcoin's value as a manifestation of entropy reduction—i.e., uncertainty minimization—within decentralized value exchange systems. Its value emerges from informational attributes, including the cryptographic order that ensures integrity, the trust embedded in the consensus mechanism, and the relational coherence within decentralized validation networks operating across the socio-technological ecosystem. To examine the plausibility of this perspective, we empirically assess how short-term shifts in Bitcoin's perceived value are predicted by variables representing values arising from the technical subsystem (Store-of-Value – SOV; Autonomy – AUT) and the social subsystem (Social-Signal Value – SSV; Hedonic-Sentiment Value – HSV). Several significant findings emerge.

The Bayesian analysis indicates that only SSV exerts a highly credible positive effect on next-day returns. This result is consistent with GITT-VT's conceptualization, which holds that Bitcoin's value is influenced by entropy regulation within the social system. When public attention intensifies (evidenced by spikes in Google Trends and media narratives), newly absorbed informational quanta enter the "buffer zone" of the collective mind, where they interact with pre-existing value quanta (e.g., legitimacy, excitement, expectations of future gains). These interactions initially increase short-term informational entropy but simultaneously trigger accelerated synthesis of insights that rapidly diffuse across the market. Such attention-driven informational inflows enhance value salience by temporarily reducing uncertainty surrounding collective narratives. The credibility of SSV provides empirical support for the proposition that Bitcoin's pricing mechanism is partly narrative-driven, with social attention operating as a short-run informational field that accelerates price movements.

Unlike the SSV, changes in SOV and AUT exhibit only moderately reliable positive effects on next-day returns. This moderate reliability can be explained through several mechanisms. First, SOV and AUT represent low-entropy informational quanta. Their underlying attributes—such as Bitcoin's scarcity mechanism, long-term holding capability, and autonomy from centralized custody—are grounded in its technological architecture rather than in fluctuating social dynamics. Autonomy-related motivations tend to reflect deeply internalized value orientations that have formed over extended temporal horizons, making them resistant to short-term shifts in public sentiment or market attention. Second, these quanta carry high informational energy, meaning they shape slow-moving structural interactions within the value formation system. Their influence is foundational, shaping long-term legitimacy, resilience, and confidence in decentralized financial logic. Thus, their internal state does not easily fluctuate in response to exogenous informational shocks, and they are therefore less likely to trigger immediate market reactions. In this sense, SOV and AUT arise from Bitcoin's technological advantages as a socio-technological infrastructure rather than from solely the dynamics of social signaling. Accordingly, the moderately reliable positive coefficients for SOV and AUT reflect the nature of these informational quanta: they are essential for maintaining and reinforcing long-term value formation, yet they seldom exert strong predictive power over short-horizon returns. Their impact is structural and foundational, not memetically contagious.

Regarding HSV, it exhibits no credible short-term effect on next-day returns. From the GITT–VT perspective, HSV primarily modulates internal subjective entropy at the individual cognitive level (e.g., emotional valuation noise). Much of HSV's short-run influence may be reflected through SSV. When attention-driven value channels dominate (e.g., narrative contagion via public discourse), HSV's explanatory power is overshadowed. Consequently, its effects become statistically ambiguous once SSV is controlled in the regression model.

Overall, the findings empirically support the GITT–VT formulation of Bitcoin as an entropy-regulating socio-technological ecosystem. Its value formation process operates through the interaction between two complementary subsystems. The low-entropy technical subsystem—characterized by protocol immutability, cryptographic security, and fixed supply—provides structural stability and anchors long-term legitimacy within the collective cognitive field. In contrast, the high-entropy socio-informational field—comprising narratives, memes, media signals, and attention waves—generates short-term volatility and memetic momentum. Our empirical model captures this duality: SOV and AUT, representing low-entropy quanta that shape long-term value trajectories but exhibit limited predictability in daily returns, whereas SSV, embodying highly interactive informational quanta, emerges as the key driver of immediate price movements. Accordingly, Bitcoin's market valuation arises from the dynamic balance between structural informational order (e.g., scarcity, autonomy) and fluctuating informational entropy (e.g., attention shocks). The observed pricing behavior can thus be seen as an emergent macroscopic outcome of multilevel informational interactions.

Also, from this perspective, Bitcoin's value may collapse under two critical conditions. First, collapse can occur if the collective consensus within the social system regarding Bitcoin's value weakens. Specifically, this would involve a decline in the societal demand for its core perceived advantages—such as its role as a store of value, the autonomy provided through decentralization, and positive expectations of future price movements. If these value expectations erode beyond the stabilizing capacity of the Bitcoin investment community, particularly due to broader shifts in socio-cultural or political value systems, the informational coherence that supports Bitcoin's long-term legitimacy may no longer be sustained. Second, value collapse may occur if Bitcoin's technical structures are compromised or fail to deliver their foundational functions, which provide informational order and facilitate low-entropy cognitive stabilization (e.g., security via cryptography, scarcity, decentralization, and immutable ledgers). While shifts in social consensus (Scenario 1) may evolve gradually or occur abruptly and often have the potential to recover if collective belief is restored, failures in the technological subsystem (Scenario 2) are typically sudden, systemic, and irreversible. A notable example of such a potential risk is the advent of quantum computing, which may undermine the cryptographic assumptions underlying Bitcoin's and Ethereum's security architectures, thereby disrupting their ability to provide structural informational stability (Chainalysis Team, 2025; Rajkumari, 2025).

This study's results highlight GITT–VT's potential broader applicability to other digital assets—particularly non-cash-flow assets whose valuation processes rely predominantly on low-entropy technical structural quanta (e.g., protocol design, cryptographic trust, decentralization) and high-entropy socio-informational quanta (e.g., narrative contagion, community discourse, attention cycles). On this basis, several research directions emerge. Comparative entropy dynamics could be examined across different classes of digital assets (e.g., NFTs, governance tokens, layer-2 scaling tokens) to evaluate how utility, network effects, and protocol-based functionality moderate the roles of structural and narrative value quanta. Entropy-based modeling may help capture how these interactions evolve across asset life cycles, or how narrative shocks may transition into structural value anchors as adoption proceeds. Future work may also integrate GITT-VT with reflexivity and market microstructure theories, develop empirical proxies and valuation metrics for informational energy and interaction potential, or explore cognitive entropy shifts among investors and online communities (Madhavan, 2000; Soros, 2013). Furthermore, longitudinal studies could investigate how upgrades in technological architecture (e.g., Taproot or Dencun enhancements) tilt value dynamics toward structural legitimacy, alongside behavioral work exploring how immersive engagement in digital ecosystems (such as metaverse platforms or crypto-based virtual environments) influences the internalization and transmission of informational quanta.

From the GITT–VT perspective, investment education in digital assets should shift from predominantly technical or speculative instruction toward cultivating informational literacy and entropy awareness. Traditional education often focuses on market indicators, technical analysis, or risk-return trade-offs, overlooking the multi-layered informational dynamics that underpin valuation. GITT–VT suggests that investors should be equipped to distinguish between low-entropy structural informational quanta (e.g., the role of cryptographic integrity, supply mechanisms, consensus protocols, decentralization) and high-entropy socio-informational quanta (e.g., attention surges, narrative shifts, memetic dynamics). Educational programs should therefore emphasize how technological foundations shape long-term intrinsic resilience, whereas narrative-driven signals may trigger short-term volatility without necessarily contributing to structural value.

By integrating GITT–VT into investment education, institutions can help learners and investors recognize that price fluctuations are emergent outcomes of interactions among multiple informational layers, rather than mere reflections of conventional financial indicators, which capture only a limited portion of the complex and dynamic infosphere. Training modules should incorporate entropy-based reasoning, helping learners evaluate whether a digital asset's momentum arises from sustained technological strength or transient attention waves. Educators could also foster cognitive resilience by teaching learners to recognize reflexivity loops, avoid emotionally driven reactions to high-frequency signals, and assess how narrative shocks may or may not evolve into long-term value anchors. Ultimately, applying GITT–VT in educational contexts encourages investors to adopt a systems-thinking approach, reducing susceptibility to memetic speculation while strengthening their capacity to evaluate digital assets based on their informational stability and evolutionary potential (Vuong & Nguyen, 2025).

While this study offers both theoretical and empirical advancements in understanding Bitcoin's informational value formation, several limitations should be acknowledged when interpreting the findings. First, the operationalization of the four value dimensions—SOV, AUT, SSV, and HSV—relies on established but necessarily simplified proxies. Indicators such as active addresses, Google Trends, and the Fear & Greed Index reflect only surface-level manifestations of deeper informational quanta within the collective cognitive field. Under the GITT–VT paradigm, value emerges from granular interactions among informational units, many of which may not be directly observable in the data streams employed here. Accordingly, these proxies likely provide only partial representations of underlying cognitive and cultural valuation processes.

Second, the study focuses on daily temporal resolution and employs a one-day lag structure, which is suitable for detecting immediate informational shocks but less effective in capturing slower, structural interactions—particularly those associated with low-entropy value quanta such as autonomy and long-term scarcity beliefs. Bitcoin's socio-technological ecosystem unfolds across multiple temporal scales; thus, dynamics evolving over weekly, monthly, or longer horizons may not be fully captured. Third, although Bitcoin's informational environment encompasses institutional signals, regulatory developments, developer community interactions, memetic propagation across social platforms, and geopolitical narratives, not all of these channels were included due to data availability and measurability constraints. Their exclusion introduces potential omitted-variable bias, particularly in modeling high-entropy informational shocks.

Finally, Bitcoin is distinctive in its transparency, memetic capacity, and ideological salience relative to both traditional financial instruments and other digital assets. Consequently, the empirical dominance of SSV observed in this study may not generalize to assets with more conventional cash-flow-based valuation mechanisms or with weaker cultural resonance. Therefore, caution is advised when extending the GITT–VT framework to other contexts without further empirical validation.